\documentclass[fleqn,usenatbib]{mnras}

\usepackage[T1]{fontenc}
\usepackage{ae,aecompl}

\usepackage{graphicx}	
\usepackage{amsmath}	
\usepackage{amssymb}	

\newcommand{\delr}{\frac{\partial}{\partial r}}

\newcommand{\mbh}{M_{\rm BH}}
\newcommand{\rsep}{R_{\rm sep}}
\newcommand{\rini}{R_{\rm ini}}
\newcommand{\rout}{r_{\rm out}}
\newcommand{\asep}{a_{\rm sep}}
\newcommand{\msun}{M_{\odot}}
\newcommand{\mdead}{m_{\rm dead}}
\newcommand{\Tdead}{T_{\rm dead}}
\newcommand{\cs}{c_{\rm s}}
\newcommand{\Mdot}{\dot M}
\newcommand{\vgw}{v_{\rm GW}}
\newcommand{\Qrad}{Q_{\rm rad}}

\title[Evolution of an Accretion Disc in BBHs]{Evolution of an Accretion Disc in Binary Black Hole Systems}

\author[S.S. Kimura, S.Z. Takahashi, \& K. Toma]{
Shigeo S. Kimura$^{1,2,3}$\thanks{E-mail: shigeo@astr.tohoku.ac.jp}, 
Sanemichi Z. Takahashi$^{2}$, 
and Kenji Toma$^{1,2}$
\\
$^1$Frontier Research Institute for Interdisciplinary Sciences, Tohoku University, Sendai 980-8578, Japan\\
$^2$Astronomical Institute, Tohoku University, Sendai 980-8578, Japan\\
$^3$Center for Particle and Gravitational Astrophysics; Department of Astronomy \& Astrophysics, Pennsylvania State University, University Park, PA 16802, USA\\
}

\date{Accepted XXX. Received YYY; in original form ZZZ}

\pubyear{2016}

\begin{document}
\label{firstpage}
\pagerange{\pageref{firstpage}--\pageref{lastpage}}
\maketitle

 \begin{abstract}
We investigate evolution of an accretion disc in binary black hole (BBH) systems and possible electromagnetic counterparts of the gravitational waves from mergers of BBHs. Perna et al. (2016) proposed a novel evolutionary scenario of an accretion disc in BBHs in which a disc eventually becomes ``dead'', i.e., the magnetorotational instability (MRI) becomes inactive. In their scenario, the dead disc survives until {\it a few seconds before} the merger event. We improve the dead disc model and propose another scenario, taking account of effects of the tidal torque from the companion and the critical ionization degree for MRI activation more carefully. We find that the mass of the dead disc is much lower than that in the Perna's scenario. When the binary separation sufficiently becomes small, the mass inflow induced by the tidal torque reactivates MRI, restarting mass accretion onto the black hole. We also find that this disc ``revival'' happens {\it more than thousands of years before} the merger. The mass accretion induced by the tidal torque increases as the separation decreases, and a relativistic jet could be launched before the merger. The emissions from these jets are too faint compared to GRBs, but detectable if the merger events happen within $\lesssim 10$ Mpc or if the masses of the black holes are as massive as $\sim 10^5 M_{\odot}$. 
\end{abstract}


\section{INTRODUCTION}\label{sec:intro}

The Laser Interferometer Gravitational-wave Observatory (LIGO) detected gravitational wave (GW) signals from merger events of binary black hole (BBH) systems \citep{LIGO16a,LIGO16d}. Although electromagnetic counterparts of GWs from mergers of BBHs were unexpected, the {\it Fermi} Gamma-ray Burst Monitor (GBM) reported detection of gamma-rays from the consistent direction of GW150914 \citep{FermiGBM16a}, which indicates possibility of a short gamma-ray burst (GRB) coincident with the merger of BBH \footnote{No electromagnetic counterparts have been reported for GW151226 \citep{Fermi16a,Pan-STAR16a}}. While some claim that the GBM event is likely to be a false signal \citep{Lyu16a,GBS16a,Xio16a}, some models are proposed to explain it \citep{PLG16a,JBC16a,Loe16a}. Possible electromagnetic counterparts in other wavelengths are also discussed \citep{MKM16a,YAO16a}. For producing powerful radiation, it is necessary to leave sufficient amount of material around the merging black holes (BHs). Here, we study disc accretion in BBHs.

The evolution of an accretion disc is determined by the efficiency of angular momentum transport. It is believed that turbulent stress induced by the magnetorotational instability (MRI) can efficiently transport the angular momentum \citep{bh91,BH98a}. This instability is active for ionized plasma, whereas it is suppressed when the ionization degree is sufficiently low \citep{Gam96a,MPH01a}. If the disc is MRI ``dead'',  the disc material can remain around BHs for a long time. \citet{PLG16a} argued that the dead disc can remain until a few seconds before the merger, and can supply energy enough to explain the GBM event. However, their model seems to ignore or misestimate a few processes that affect evolution of an accretion disc in binary systems. One important process is the effect of tidal torque, which prevents the disc material from expanding outward beyond the tidal truncation radius \citep{IO94a}. This causes the disc mass to decrease faster than that of the well-known self-similar solutions \citep[e.g.][]{CLG90a,MPH01a,PLG16a}. The tidal torque also causes to heat up the outer edge of the dead disc in the late phase of evolution, which can eventually reactivates MRI. Another important point is the critical ionization degree for MRI activation. MRI is usually active for very low ionization degree \citep[e.g.][]{Gam96a,SM99a}, and the critical temperature for MRI activation is very low, typically less than a few thousands K. This causes the MRI activation tens of thousands years before the merger.

\begin{figure}
   \includegraphics[width=\linewidth]{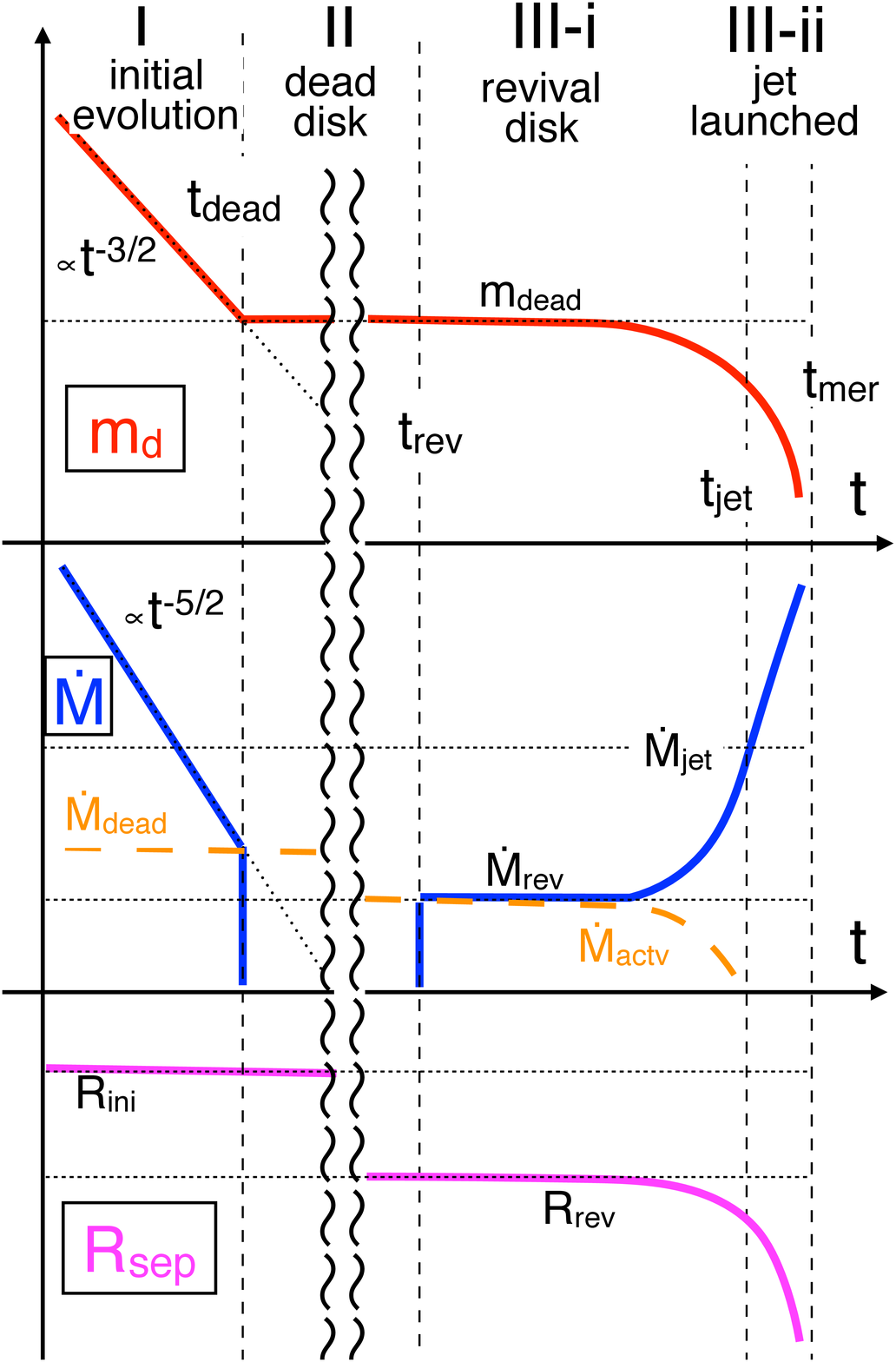}
    \caption{Schematic evolutionary tracks of the disc mass (red), the mass accretion rate (blue), and the binary separation (magenta). Note that this is double logarithmic plot and that phase II is much longer than the other phases.}
    \label{fig:evolution}
 \end{figure}

In this paper, we improve the dead disc model and propose another scenario, which predicts electromagnetic counterparts of GWs whose luminosity increases with time. In Figure \ref{fig:evolution}, we show the schematic evolutionary tracks of the disc mass $m_{\rm d}$, the mass accretion rate $\dot M$, and the binary separation $R_{\rm sep}$. The disc experiences three phases. At first, the disc forgets its initial condition through viscous evolution. Then, the disc mass and the accretion rate decrease with radiative cooling, which leads to decrease of the ionization degree (phase I). This eventually suppresses MRI, forming a dead disc that remains around the BH until the binary separation sufficiently decreases (phase II). Then, the heating by the tidal torque from the companion becomes effective, which reactivates MRI in the entire region of the disc, restarting accretion onto the BH (phase III-i). This disc ``revival'' happens many years before the merger \footnote{\citet{PLG16a} mentioned a low-luminosity and long-lasting transient preceding the merger by the MRI reactivation due to photons from the outer rim, although they did not discuss it in detail.}. We describe this model in detail in Section \ref{sec:evolv}. The mass accretion rate increases as the separation decreases, and a relativistic jet could be launched owing to high accretion rate (phase III-ii). We estimate flux of electromagnetic emission from the jet and discuss its detectability in Section \ref{sec:detect}. Section \ref{sec:summary} is devoted to summary and discussion.

\section{Evolution of a disc in BBH systems}\label{sec:evolv}
\subsection{Initial evolution}\label{sec:initial}

We consider an equal-mass binary of initial separation $\rini$ and mass of BHs $\mbh$, where the separation should be small such that the binary can merge in the Hubble time. Some mechanisms are proposed to realize this situation, such as the common envelope evolution \citep{KIH14a,BHB16a} and/or the friction by dense gas \citep{BKH16a}. We focus on an accretion disc around one of the BHs. We do not discuss the origin of this disc, which might be fallback material of supernova explosion \citep[e.g.,][]{PDC14a} or a tidally disrupted object \citep[e.g.,][]{SM11a}.

Consider a gas ring around a BH. The ring expands both inward and outward due to the viscous diffusion to become an accretion disc \citep[e.g.,][]{pri81}. When the outer radius of the disc, $\rout$, becomes close to $\rini$, the tidal torque from the companion prevents the disc from expanding outward \citep{PP77a,AL94a,IO94a}. The balance between the viscous torque and tidal torque determines the disc radius, and it is expected that the outer radius of the disc is fixed at $\rout\sim\asep\rini$, where we introduce a separation parameter $\asep$. We fix $\asep=0.3$ in this paper for simplicity \citep{Pac77a}. The disc expands to $\rout$ in the viscous time \citep[e.g.][]{pri81}
\begin{eqnarray}
 t_{\rm vis}&=&{1\over\alpha\Omega_{\rm K}}\left(\frac{\rout}{H}\right)^2\nonumber\\
 &\sim&2.6\times10^{4}a_{-0.5}^{3/2}R_{\rm i,12}^{3/2}M_{1.5}^{-1/2}\alpha_{-1}^{-1}\left(\frac{\rout}{H}\right)^2\rm~s,\label{eq:tvis}
\end{eqnarray}
where $\Omega_{\rm K}=\sqrt{G\mbh/r^3}$ is the Keplerian angular velocity, $H=\cs/\Omega_{\rm K}$ is the scale height ($\cs$ is the sound speed), $M_{1.5}=\mbh/30M_{\odot}$, $\alpha_{-1}=\alpha/0.1$, $R_{\rm i,12}=\rini/(10^{12}{\rm~cm})$, and $a_{-0.5}=\asep/0.3$. We use the alpha prescription for viscosity, $\nu=\alpha\cs^2/\Omega_{\rm K}$. On the other hand, the time scale of GW inspiral is \citep[e.g.][]{ST83a}
\begin{equation}
t_{\rm mer}={5\over512}\frac{c^5}{G^3}{\rini^4\over\mbh^3}\sim3.8\times10^{15}R_{\rm i,12}^4M_{1.5}^{-3}\rm~s.
\end{equation}
We can see $t_{\rm vis}<t_{\rm mer}$ for $\rout/H\lesssim10^{5}$, which is valid in all the situations we usually expect.
Thus, the disc forgets its initial mass and/or radius due to viscous evolution before the merger.

For the well-known solution of an accretion disc around single BHs, the disc outer radius increases with time as a result of the outward angular momentum transport \citep{LP74a}. On the other hand, in a binary system, the angular momentum of the disc material is carried to the companion by the tidal torque. Therefore, the disc material can accrete onto the BH without increasing the disc outer radius. Note that the tidal heating and torque are effective only in very thin outer rim located just outside $r_{\rm out}$ \citep{IO94a}. Almost all the mass is in the viscously heated region of $r\le r_{\rm out}$, and the mass that expands beyond $\rout$ is expected to be negligible. Note that the merging time $t_{\rm mer}$ is unchanged by the angular momentum transport from the disc to the companion if the mass of the disc is much lower than that of the companion.

We consider evolution of the disc in a binary system, assuming opacity of the disc is constant, $\kappa=0.4\rm~cm^2~g^{-1}$, for simplicity. This treatment is not accurate very much because opacity is a function of temperature and density for $T\lesssim 10^5$ K \citep[e.g.][]{CW84a,BL94a,ZHG09a}. However, this treatment enables us to make a fully analytic calculation with an acceptable accuracy. While the value of $\kappa$ can vary an order of magnitude, the temperature only changes factor of a few owing to $T\propto \kappa^{1/4}$. 
 The viscous heating and radiative cooling rates are
\begin{eqnarray}
 Q_{\rm vis}&=&\frac{9}{8}\nu\Sigma\Omega_{\rm K}^2,\label{eq:q+}\\
 \Qrad&=&{8\sigma_{\rm sb}T^4\over3\kappa\Sigma},\label{eq:q-}
\end{eqnarray}
respectively. The thermal balance, $Q_{\rm vis}=\Qrad$, gives the disc temperature as 
\begin{equation}
 T=\left({27k_{\rm B}\kappa\over64\sigma_{\rm sb}m_p}\right)^{1/3}\alpha^{1/3}\Omega_{\rm K}^{1/3}\Sigma^{2/3}, \label{eq:temperature}
\end{equation}
where we use $\cs^2=k_{\rm B}T/m_p$. The viscous time is shorter in the inner region of the disc, where the steady state is realized \citep[e.g.,][]{LP74a}. The mass accretion rate onto the BH is estimated to be 
\begin{equation}
 \Mdot = 3\pi \nu \Sigma \propto \Sigma^{5/3}\Omega_{\rm K}^{-2/3}.\label{eq:mdot}
\end{equation}
 Since this mass accretion rate is constant for the inner region, the radial profile of the surface density is $\Sigma\propto r^{-3/5}$. Using this profile, we estimate the disc mass to be
\begin{equation}
 m_{\rm d}=\int_{r_{\rm in}}^{\rout}2\pi\Sigma rdr\approx{10\pi\over7}\Sigma_{\rm out}\rout^2,\label{eq:msigma}
\end{equation}
where $\Sigma_{\rm out}=\Sigma(r=\rout)$. Note that treatment of $\rout$ (fixed as $\asep\rsep$) is crucial for tracking the evolution of $m_{\rm d}$ because it strongly depends on $\rout$. 

Ignoring wind mass loss, the disc mass decreases according to Equation (\ref{eq:mdot}). Then, we can write the evolution of disc mass as
\begin{equation}
 \frac{dm_{\rm d}}{dt}=-\Mdot = f(\rini,~\asep,~\mbh,~\alpha)m_{\rm d}^{5/3},
\end{equation}
where we set $\rsep\approx \rini$, since $t_{\rm vis}\ll t_{\rm mer}$. Then, we can integrate this equation and obtain
\begin{equation}
 m_{\rm d} = m_{\rm d,ini}\left({t\over t_{\rm ini}}\right)^{-3/2},\label{eq:md}
\end{equation}
where $m_{\rm d,ini}$ is the disc mass at the time $t=t_{\rm ini}$. 

 \begin{figure}
   \includegraphics[width=\linewidth]{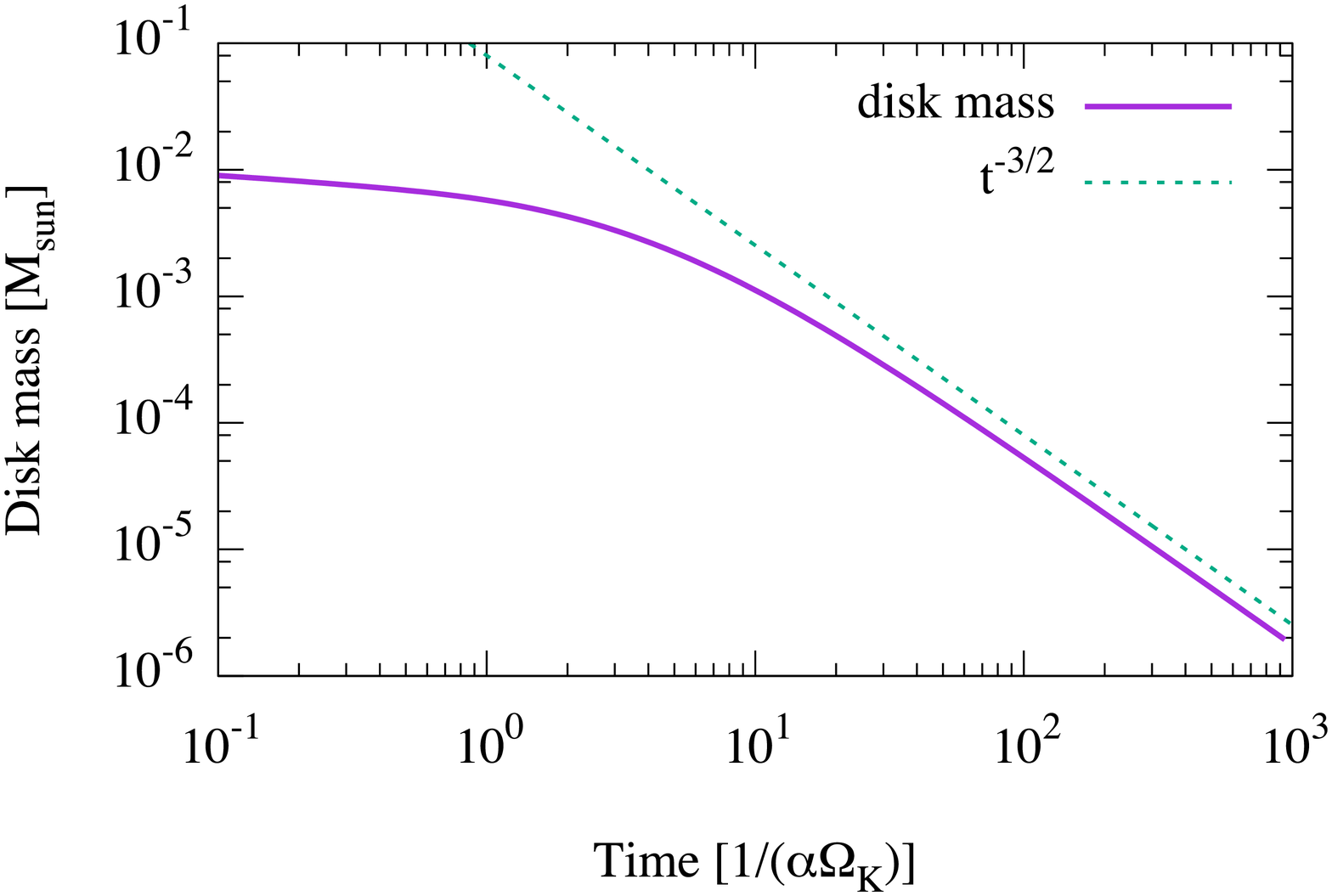}
   \includegraphics[width=\linewidth]{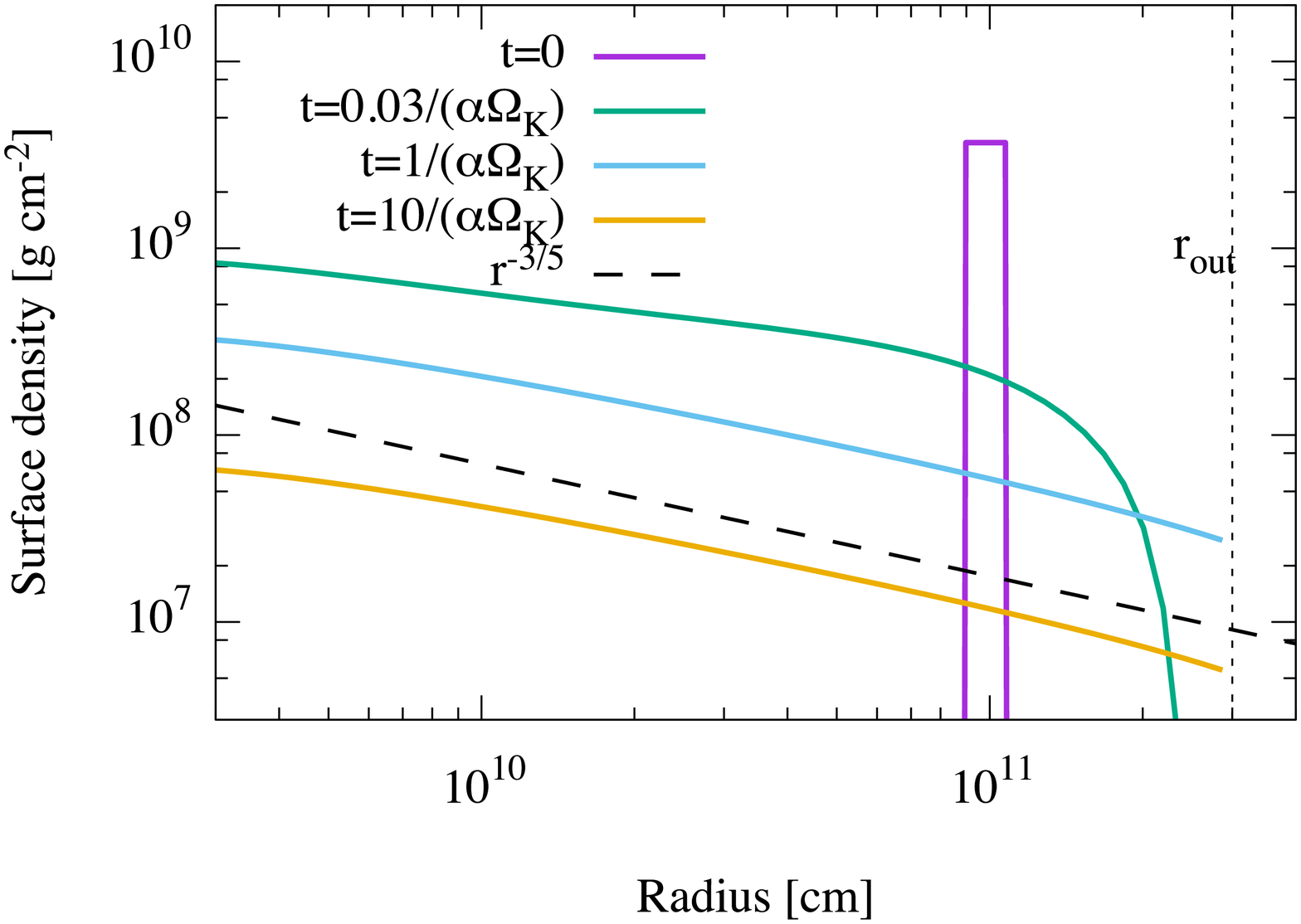}
    \caption{The results of the numerical calculation of the diffusion equation. The upper panel shows the evolution of disc mass. The numerical calculation (solid line) matches the analytic model (dotted line) for $t\gtrsim 5/(\alpha\Omega_{\rm K})$. The lower panel shows the radial profile of the surface density. The vertical dotted line shows the outer boundary. The profiles are single power-law for $t\gtrsim 1/(\alpha\Omega_{\rm K})$, and the material does not accumulate near $r_{\rm out}$.}
    \label{fig:disc}
 \end{figure}

To confirm this scaling relation, we numerically solve the diffusion equation of viscous disc evolution:
\begin{equation}
 \frac{\partial \Sigma}{\partial t} = \frac{1}{r}\delr\left[\frac{1}{dj/dr}\delr \left(\nu \Sigma r^3 \frac{d\Omega}{dr}\right)\right],\label{eq:diffusion}
\end{equation}
with a boundary condition $\dot M=0$ at $r=r_{\rm out}$. This treatment corresponds to the assumption (introduced above) that the tidal torque is effective only in the very thin outer rim just outside $r=r_{\rm out}$ \citep{IO94a}. That is, the disc evolution at $r<r_{\rm out}$ is governed by the viscous torque as described by Equation (\ref{eq:diffusion}), and the tidal torque is balanced to the viscous torque just at $r= r_{\rm out}$. The disc material at $r=r_{\rm out}$, receives the angular momentum from the material at $r < r_{\rm out}$ by the viscous torque. The same amount of angular momentum is transported to the companion by the tidal torque, which makes the angular momentum flux constant at $r=r_{\rm out}$. Therefore, the disc material at $r=r_{\rm out}$ does not expand further. We initially put a gas ring of $m_{\rm d,ini}=0.01\msun$ at $r=10^{11}$ cm. We use the reference parameter set ($\alpha=0.1$, $\mbh=30\msun$, $\rini=10^{12}$ cm, and $\asep=0.3$). We show the results of the numerical calculation in Figure \ref{fig:disc}. The upper panel shows the evolution of disc mass, which matches our analytic model in later time. The lower panel shows the radial profiles of the surface density at $t=0,~t=0.03/(\alpha\Omega_{\rm K})$, $t=1/(\alpha\Omega_{\rm K})$, and $t=10/(\alpha\Omega_{\rm K})$, where we use $\Omega_{\rm K}$ at $r=r_{\rm out}$. The disc material expands both inward and outward, and reaches the outer boundary at $t\sim 1/(\alpha\Omega_{\rm K})$. After that, its profile are expressed as a single power-law of $r^{-3/5}$. This means that the disc material does not accumulate near $r=r_{\rm out}$, implying that the mass may be estimated by Equation (\ref{eq:msigma}).

We have considered the thin and gas-pressure dominant disc. This requires $H/r<1$ and $p_{\rm rad}/p_{\rm gas}<1$, where $p_{\rm gas}=\Sigma c_{\rm s}^2/(2H)$ and $p_{\rm rad}=aT^4$ ($a$ is the radiative constant). Our calculation in Figure \ref{fig:disc} is inconsistent with this treatment in the early phase where the disc is very hot due to its high accretion rate. We obtain $H/r\simeq 1.3$ and $p_{\rm rad}/p_{\rm gas}\simeq 3.4\times10^6$ for the initial ring of $t=0$. However, these inconsistency does not affect the following discussion (see Section \ref{sec:summary} for detail).

If MRI is always active, this scaling relation is applicable all the time. Setting $t_{\rm ini} = t_{\rm vis}$, we estimate the disc mass at the time of merger and find that it is $10^{-11}$ times lower than the initial disc mass, where we use the reference parameter set used in Equation (\ref{eq:tvis}) and $(\rout/H)\sim 100$. Even if $m_{\rm d,ini}$ is as massive as $100 \msun$, the disc mass of the merging time is $10^{-9}~\msun$. This is too low to produce energetic electromagnetic counterparts of GW signals, which requires $m_{\rm d}\sim 10^{-5}M_{\odot}$ \citep{MKM16a,Lyu16a}.

\subsection{Formation of a dead disc}\label{sec:dead}
 
The disc cools down as the disc becomes lighter, which changes the disc state from fully-ionized plasma to almost neutral. The MRI is inactive if the ionization degree sufficiently decreases. The condition for MRI to be active is \citep[e.g.,][]{SM99a,OH11a,FOT14a}
 \begin{equation}
  \Lambda={v_{\rm A}^2\over\eta\Omega_{\rm K}}>1,
 \end{equation}
where $\Lambda$ is the Elsasser number, $v_{\rm A}$ is the Alfven velocity, and $\eta$ is the resistivity. The resistivity in accretion discs, where the Ohmic dissipation is dominant, is $\eta=234(T/1{\rm K})^{1/2}\chi_e^{-1}\rm cm^2 s^{-1}$ \citep{BB94a}. Writing $v_{\rm A}^2=2\cs^2/\beta_{\rm pl}$, the instability condition at the disk outer radius $\rout\simeq \asep\rsep$ is
 \begin{eqnarray}
  \chi_e > \chi_{\rm dead} = {117T^{1/2}\beta_{\rm pl}\Omega_{\rm K}\over\cs^2}\nonumber\\
 \simeq9.9\times10^{-10}\beta_2T_{3.5}^{-1/2}M_{1.5}^{1/2}a_{-0.5}^{-3/2}R_{\rm s,12}^{-3/2},\label{eq:chi_cr}
\end{eqnarray}
where $\chi_e=n_e/n$ is the ionization degree ($n=\Sigma/(2m_pH)$ is the total number density), $\beta_{\rm pl}=8\pi P/B^2$ is the plasma beta ($P$ is the gas pressure), $T_{3.5}=T/(3000\rm~K)$, $\beta_2=\beta_{\rm pl}/(10^2)$, and $R_{\rm s,12}=R_{\rm sep}/(10^{12}\rm~cm)$. The MRI is active even for such a low ionization degree. 

We calculate the ionization degree in the accretion disc by solving the Saha's equation. Since BHs heavier than $\sim10M_{\odot}$ are expected to form only under low-metalicity environments \citep{LIGO16b}, we consider pure hydrogen discs. Then, the Saha's equation is
\begin{equation}
{\chi_e^2\over1-\chi_e}=\frac{1}{n}\left(\frac{2\pi m_ek_{\rm B}T}{h^2}\right)^{3/2}\exp\left(-\frac{E_{\rm i}}{k_{\rm B}T}\right), \label{eq:saha}
\end{equation} 
where $E_{\rm i}=13.6$ eV. Since the ionization degree exponentially decreases with temperature, the outer edge of the viscously heated region ($r=\rout$) is the first place which becomes dead. When the dead region appears at $r=\rout$, the mass inflow to the inner region ($r<\rout$) stops, which causes the inner region of $r<\rout$ to cool down rapidly due to the lack of heating source. Thus, the dead region propagates inward, and the entire region of the disc becomes dead (formation of a dead disc). Using equation (\ref{eq:temperature}), (\ref{eq:chi_cr}), and (\ref{eq:saha}), we calculate the critical temperature $\Tdead$ below which MRI is dead for given $\beta_{\rm pl}$ and $\rini$. We find that  $2600{\rm~K}<\Tdead<4000$ K for $10\le\beta_{\rm pl}\le10^3$ and $10^{11}{\rm~cm}<\rini<3\times10^{12}$ cm. The parameter dependence of $\Tdead$ is so weak that we can hereafter estimate physical quantities by approximating $\Tdead\sim3000$ K. 

We should note that the disc temperature does not continuously approach $T_{\rm dead}$ from a higher temperature. Instead, the disc temperature rapidly drops due to the thermal instability induced by the strong temperature dependence of opacity \citep{LFP85a,Can93a,Las01a}. The critical temperature for the thermal instability is represented as \citep[Equation A.8 in][]{Las01a} 
\begin{equation}
T_{\rm TI}\simeq 3.9\times10^4 \alpha_{-1}^{-0.21} M_{1.5}^{-0.02} a_{-0.5}^{0.05}R_{\rm s,12}^{0.05} \rm~K.
\end{equation}
Once the temperature becomes lower than $T_{\rm TI}$ at the outer edge of the viscously heated region, the disc temperature immediately drops to $T\lesssim 3000$ K because there is no stable solution between $T_{\rm TI}$ and $\sim 3000$ K. After the thermal instability takes place at the outer edge, a cooling wave propagates inward, and the entire region of the disc changes to a cold state $T\sim$ 3000 K \citep{Can93a,Las01a}, forming a dead disc. 

Taking this thermal instability into account, the mass of the dead disc is estimated to be 
\begin{eqnarray}
 \mdead\approx{80\pi\over21}\left({\sigma_{\rm sb}m_{\rm p}\over3k_{\rm B}\kappa}\right)^{1/2}\left({T_{\rm TI}^3\over\alpha\Omega_{\rm K}}\right)^{1/2}\rout^2 \label{eq:mdead}\\
 \simeq5.1\times10^{-7}T_{\rm T,4.6}^{3/2}\alpha_{-1}^{-1/2}M_{1.5}^{-1/4}R_{\rm s,12}^{11/4}a_{-0.5}^{11/4}M_{\odot}\nonumber,
\end{eqnarray}
where $T_{\rm T,4.6}=T_{\rm TI}/(3.9\times10^4~\rm K)$ and we use Equation (\ref{eq:temperature}) and (\ref{eq:msigma}). Note that this $m_{\rm dead}$ is independent of the disc initial condition, although the initial mass $m_{\rm d,ini}$ appears in Equation (\ref{eq:md}). The initial condition affects the time when the disc becomes dead, $t_{\rm dead}$. For example, $t_{\rm dead}\sim2$ yr for the fiducial parameter set shown in Figure \ref{fig:disc}. We can see $t_{\rm dead} \propto m_{\rm d,ini}^{2/3}$ from Equation (\ref{eq:md}). Thus, $t_{\rm dead}$ is much shorter than the binary evolution time, $t_{\rm GW}$, for usual situation, which means $R_{\rm sep}\simeq R_{\rm ini}$ is satisfied when the dead disc forms. We find that for our fiducial parameter set, $\mdead$ is lower than the required mass for the luminous electromagnetic counterparts, $\sim 10^{-5}~\msun$ \citep{MKM16a,Lyu16a}. Since $\mdead$ strongly depends on $\rini$, the dead disc for $\rini\gtrsim10^{13}$ cm can be massive enough to emit luminous electromagnetic counterparts. While $t_{\rm mer}$ is longer than Hubble time for such a wide separation,  rapid separation decrease might occur by some mechanisms, such as friction by dense gas \citep[e.g.,][]{BKH16a}. The critical mass accretion rate for MRI stabilization is 
\begin{eqnarray}
 \Mdot_{\rm dead}&=&3\pi\nu\Sigma_{\rm out}\nonumber\\
&=&8\pi\left({k_{\rm B}\sigma_{\rm sb}\over 3\kappa m_p}\right)^{1/2}\alpha^{1/2}\Omega_{\rm K}^{-3/2}T_{\rm TI}^{5/2}\label{eq:mdotdead}\\
 &\simeq&2.0\times10^{19}\alpha_{-1}^{1/2}M_{1.5}^{-3/4}R_{\rm s,12}^{9/4}a_{-0.5}^{9/4}T_{T,4.6}^{5/2}\rm~g~s^{-1}\nonumber,
\end{eqnarray}
where we use Equations (\ref{eq:temperature}) and (\ref{eq:mdot}). 
Note that this value is several times higher than that in \cite{Las01a} mainly due to our simple treatment of the opacity.
Once the disc becomes dead, it remains until the binary separation sufficiently decreases.

\subsection{Revival of a dead disc}\label{sec:revival}

The binary separation, $\rsep$, decreases owing to emission of gravitational waves even for the dead disc phase (phase II in Figure \ref{fig:evolution}). The decrease of the binary separation causes the decrease of the radius of the dead disk, $r_{\rm out}$, beyond which the tidal torque is effective. Then, the amount of gas in the outer rim ($r>r_{\rm out}$) increases. The angular momentum of the gas in $r>r_{\rm out}$ is transported to the companion by the tidal torque. This induces the mass inflow from the outer rim to the dead disc. Therefore, the decrease of the binary separation provides the mass inflow from the outer rim to the dead disc, which can reactivate the MRI. For the standard discs, the critical accretion rate for MRI activation for the arbitrary radius $r$ is estimated to be 
\begin{eqnarray}
 \Mdot_{\rm actv}(r)&=&3\pi\nu\Sigma \nonumber\\
&=&8\pi\left({k_{\rm B}\sigma_{\rm sb}\over 3\kappa m_p}\right)^{1/2}\alpha^{1/2}\Omega_{\rm K}^{-3/2}T_{\rm dead}^{5/2}\label{eq:mdotact}\\
 &\simeq&2.7\times10^{15}\alpha_{-1}^{1/2}M_{1.5}^{-3/4}r_{11}^{9/4}T_{d,3.5}^{5/2}\rm~g~s^{-1}\nonumber,
\end{eqnarray}
where $r_{11}=r/(10^{11}\rm~cm)$ and we again use Equations (\ref{eq:temperature}) and (\ref{eq:mdot}). Using Equations (\ref{eq:q+}), (\ref{eq:q-}), and (\ref{eq:mdotact}), we can obtain the relation between $\Mdot_{\rm actv}$ and $\Tdead$ as 
\begin{equation}
 \frac{3}{8\pi}\Mdot_{\rm actv}\Omega_{\rm K}^2\sim {8\sigma_{\rm sb}\Tdead^4\over3\kappa\Sigma}. \label{eq:mdotTdead} 
\end{equation}
Note that the thermal instability does not affect $\dot M_{\rm actv}$ because the solution of the lower branch has a stable solution up to $T\sim 1.2\times10^4$ K $>\Tdead$ \citep[Equation A.7 in][]{Las01a}.  Note also that $\dot M_{\rm actv}$ is the increasing function of the radius $r$. In this situation, the inner region is always active for MRI whenever the outer region is active as discussed below.

The decreasing rate of the separation is \citep[e.g.][]{ST83a}
\begin{equation}
 \vgw=\frac{d\rsep}{dt}=-\frac{128G^3\mbh^3}{5c^5\rsep^3}. \label{eq:vGW}
\end{equation}
Assuming $\rout=\asep \rsep$ with constant $\asep$, the decreasing rate of the disc outer radius is written as $\asep\vgw$. We write the surface density of the dead disc as $\widetilde{\Sigma}\sim\mdead/(\pi\rout^2)$. Then, the mass inflow rate caused by the separation decrease is estimated to be
\begin{equation}
 \Mdot_{\rm SD}=-2\pi\rout\widetilde{\Sigma}\asep\vgw\sim-{2\mdead\vgw\over\rsep}.
\end{equation}
This mass inflow releases the gravitational energy, causing to heat up gas in the outermost region. The heating rate by the mass inflow can be represented as $\sim \eta_{\rm g}\Mdot_{\rm SD}\Omega_{\rm K}^2$ \citep{KFM08a}, where $\eta_{\rm g}$ is the heating efficiency of released gravitational energy. The temperature at the outer most region is determined by 
\begin{equation}
 \eta_{\rm g}\Mdot_{\rm SD}\Omega_{\rm K}^2\sim {8\sigma_{\rm sb}T^4\over3\kappa\Sigma},\label{eq:mdotheating}
\end{equation}  
Since Equations (\ref{eq:mdotTdead}) and (\ref{eq:mdotheating}) are the same form if we assume $\eta_{\rm g}=3/(8\pi)$ \footnote{the dependence of physical quantities, such as $R_{\rm rev}$, on $\eta_{\rm g}$ is very weak, and it does not affect our conclusion.}, the condition $T_{\rm dead} < T$ is identical to $\dot M_{\rm SD} >\dot M_{\rm actv}$. When $\Mdot_{\rm SD}$ becomes higher than $\Mdot_{\rm dead}$ at $\rout$, the outer edge of the dead disc becomes MRI active. The gas in the outermost MRI active region inevitably falls to the inner region of $r<r_{\rm out}$ even if the inner region is dead, because the viscous stress transports the angular momentum at the active region \citep[e.g.][]{ZHG10a,SMI10a}. This heats up gas at the inner dead region with the heating rate $\sim \eta_{\rm g}\Mdot_{\rm SD}\Omega_{\rm K}^2$. This heating rate is high enough to activate MRI in the inner dead region because $\Mdot_{\rm SD}>\Mdot_{\rm actv}(\rout)>\Mdot_{\rm actv}(r)$, and the MRI active region propagates inward with the local viscous time. Therefore, once the mass inflow activates MRI at $r\sim \rout$, the whole part of the disc inevitably becomes active, restarting the mass accretion onto the BH. This disc ``revival'' happens when $\Mdot_{\rm SD}=\Mdot_{\rm actv}(\rout)$. The separation at that time is 
\begin{eqnarray}
 R_{\rm rev}=\left[{32\mdead\over5\pi c^5}\left(\frac{3m_p\kappa}{k_{\rm B}\sigma_{\rm sb}\alpha}\right)^{1/2}{(G\mbh)^{15/4}\over\asep^{9/4}\Tdead^{5/2}}\right]^{4/25} \\
 \sim1.4\times10^{11}m_{-6.3}^{4/25}\alpha_{-1}^{-2/25}a_{-0.5}^{-9/25}M_{1.5}^{3/5}T_{d,3.5}^{-2/5}\rm~cm\nonumber,
\end{eqnarray}
where $m_{-6.3}=\mdead/(5\times10^{-7}M_{\odot})$. When the disc revives, $t_{\rm vis}\sim t_{\rm GW}$ is satisfied because $\Mdot_{\rm dead}\sim\mdead/t_{\rm vis}$ and $\Mdot_{\rm SD}\sim\mdead/t_{\rm GW}$.

After the revival, the separation decreasing rate is likely to control the mass accretion rate onto the BH as \footnote{Two $\dot M$ introduced in this subsection is different: $\dot M_{\rm SD}$ is the mass inflow rate from the outer rim to the dead disc and $\dot M_{\rm GW}$ is the mass accretion rate from the revival disc to the central BH.}
\begin{equation}
 \Mdot_{\rm GW}=-2\pi\rout\Sigma_{\rm out}\asep\vgw=-{7m_{\rm d}\vgw\over5\rsep},\label{eq:mdotGW}
\end{equation}
where we use the disc profile of the steady disc solution as phase I, $\Sigma\propto r^{-3/5}$ and $\Sigma_{\rm out}=7m_{\rm d}/(10\pi\rout^2)$ (see Subsection \ref{sec:initial}). The disc temperature is determined so that $t_{\rm GW}\sim t_{\rm vis}$ at $r=r_{\rm out}$ is satisfied. 
Using the relation $dm_{\rm d}/dt=-\Mdot_{\rm GW}$, we can write $dm_{\rm d}/d\rsep=7m_{\rm d}/(5\rsep)$, which is integrated as
\begin{equation}
 m_{\rm d}=\mdead\left(\frac{\rsep}{R_{\rm rev}}\right)^{7/5}.\label{eq:mdRsep}
\end{equation}
Since the mass of the dead disc is conserved during phase II, $m_{\rm d}=\mdead$ for $\rsep=R_{\rm rev}$. Also, we can integrate Equation (\ref{eq:vGW}) and obtain
\begin{equation}
 \rsep=R_{\rm rev}\left(\frac{t_{\rm mer}-t}{t_{\rm mer}-t_{\rm rev}}\right)^{1/4},\label{eq:Rsept}
\end{equation}
where $t_{\rm rev}$ is the time when the disc revives. The time from the revival to the merger is very long,
\begin{eqnarray}
 t_{\rm mer}-t_{\rm rev}&=&{5\over512}{c^5\over G^3}{R_{\rm rev}^4\over\mbh^3} \\
&\sim&1.3\times10^{12}m_{-6.3}^{16/25}\alpha_{-1}^{-8/25}a_{-0.5}^{-36/25}M_{1.5}^{-3/5}T_{d,3.5}^{-8/5}\rm~s.\nonumber
\end{eqnarray}
Equation (\ref{eq:mdRsep}) and (\ref{eq:Rsept}) lead to 
\begin{equation}
 m_{\rm d}=\mdead\left(\frac{t_{\rm mer}-t}{t_{\rm mer}-t_{\rm rev}}\right)^{7/20}.
\end{equation}
The mass accretion rate is 
\begin{equation}
 \Mdot_{\rm GW}={7\mdead\over20(t_{\rm mer}-t_{\rm rev})}\left(\frac{t_{\rm mer}-t}{t_{\rm mer}-t_{\rm rev}}\right)^{-13/20}.
\end{equation}
For $t<t_{\rm mer}$, $\Mdot_{\rm GW}$ is almost constant,
\begin{eqnarray}
\Mdot_{\rm rev}&\simeq&{7\mdead\over20(t_{\rm mer}-t_{\rm rev})} \\
&\sim&2.6\times10^{14}m_{-6.3}^{9/25}\alpha_{-1}^{8/25}a_{-0.5}^{36/25}M_{1.5}^{3/5}T_{d,3.5}^{8/5}\rm ~g~s^{-1}\nonumber. 
\end{eqnarray}
This solution indicates that the tidal torque controls the mass accretion rate such that $\Mdot_{\rm rev}\sim\Mdot_{\rm actv}$, marginally keeping steady accretion, as shown in Figure \ref{fig:evolution}. This accretion rate is so low that it is difficult to observe it. For $t\lesssim t_{\rm mer}$, the mass accretion rate increases with time as $\propto(t_{\rm mer}-t)^{-13/20}$. This situation continues until $\alpha^{-1}\Omega_{\rm K}^{-1} > \rsep/\vgw$ is satisfied, which is just before the merger (0.005 s for $\mbh=30M_{\odot}$ and $\alpha=0.1$). After that, a smooth accretion flow no longer exists, and a shocked-violent accretion is likely to take place \citep{FDM15a}.

\section{Detectability of electromagnetic counterparts}\label{sec:detect}

\begin{table*}
\begin{center}
\caption{Parameters and physical quantities related to the electromagnetic counterparts from jets \label{tab:EM}}
\begin{tabular}{|c|cc|cccc|}
\hline
model & $\mbh[M_{\odot}]$ & $\rini$ [cm]& ($t_{\rm mer}-t_{\rm jet}$) [s] & $L_{\rm AG}$ [erg s$^{-1}$] & $T_{\rm AG}$ [s] & $d_{\rm L,limit}$ [Mpc]\\
\hline
A & 30 & $3\times10^{12}$ & 3.0$\times10^5$ & 3.8$\times10^{40}$ & 1.5$\times10^3$ & 19\\
B & 10$^3$ & $3\times10^{13}$ & 4.7$\times10^6$ & 1.3$\times10^{42}$ & 1.0$\times10^4$ & 1.1$\times10^2$\\
C & 10$^5$ & $10^{15}$ & 4.36$\times10^8$ & 1.3$\times10^{44}$ & 2.1$\times10^5$ & 1.1$\times10^3$\\
\hline
\end{tabular}
\end{center}
\end{table*}

  \begin{figure}
   \includegraphics[width=\linewidth]{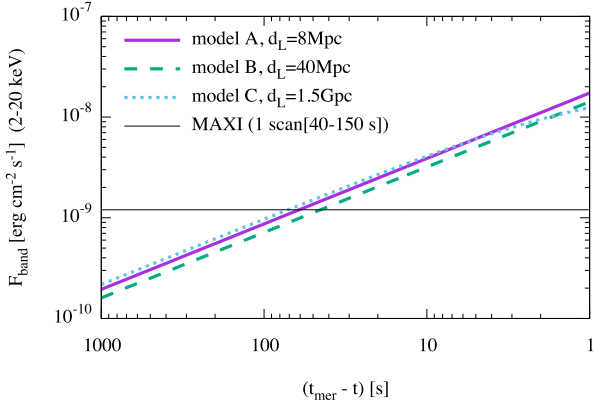}
    \caption{The photon fluxes from the internal shocks for models A, B, and C. The 1-scan sensitivity of MAXI GSC is also plotted. It is found that all the models are detectable by MAXI GSC. }
    \label{fig:detect}
  \end{figure}

The accretion rate can exceed the Eddington accretion rate (at $t_{\rm jet}$ in Figure \ref{fig:evolution}). A relativistic jet is expected to be launched from the accreting BH in such a situation \citep{TNM11a}. We consider $\Mdot_{\rm GW}\sim\Mdot_{\rm jet}\equiv10L_{\rm Edd}/c^2$ at the jet launching. We estimate the kinetic luminosity of the jet to be $L_{\rm jet}\sim \Mdot_{\rm GW}c^2$. The high-energy photons are produced in internal shocks within the jet, whose bolometric luminosity and flux at the Earth are estimated to be $L_\gamma\sim\eta_\gamma L_{\rm jet}$ and  $F_\gamma=L_\gamma/(4\pi d_{\rm L}^2)$, respectively, where $\eta_\gamma$ is the radiative efficiency of the internal shocks and $d_{\rm L}$ is the luminosity distance. The observed time of these photons after the jet launch is $\Delta t_\gamma\sim R_{\rm s}/c\sim3\times10^{-4}M_{1.5}$ s, where $R_{\rm s}$ is the Schwarzschild radius of the BH, so that these photons arrive at the Earth before the GW signal. 

The jet sweeps up gas surrounding the BBH and creates an external shock, which emits broadband photons, i.e., afterglow. The bolometric luminosity of the afterglow is estimated to be
\begin{equation}
 L_{\rm AG}\sim{E_{\rm jet}(t)\over t-t_{\rm jet}}=\frac{\int_{t_{\rm jet}}^t L_{\rm jet}dt'}{t-t_{\rm jet}} ,
\end{equation}
where $E_{\rm jet}(t)$ is the time integrated energy and $t_{\rm jet}$ is the jet launching time. Since the jet luminosity for $t_{\rm jet}<t<t_{\rm mer}$ is almost constant \footnote{For $t_{\rm jet}<t<t_{\rm mer}$, $\Mdot_{\rm GW}$ appears to be rapidly increasing in Figure \ref{fig:evolution}, while we can see $\Mdot_{\rm GW}\approx$ const if we plot $\Mdot_{\rm GW}$ as a function of $(t-t_{\rm jet})$.},  $E_{\rm jet}(t)$ is proportional to $(t-t_{\rm jet})$. Thus, this luminosity is almost constant, $L_{\rm AG}\sim\Mdot_{\rm GW}c^2\sim10L_{\rm Edd}$ for $t_{\rm jet}<t<t_{\rm mer}$. The photons of the afterglow arrive at the Earth both before and after the GW signal. The duration of the bright afterglow phase after the GW signal is
\begin{equation}
 T_{\rm AG}\sim{R_{\rm AG}\over2c\Gamma_{\rm jet}^2}=\left({3E_{\rm jet}(t=t_{\rm mer})\over4\pi m_pc^5n_{\rm ext}\Gamma_{\rm jet}^8}\right)^{1/3},
\end{equation}
where $R_{\rm AG}$ is the deceleration radius, $n_{\rm ext}$ is the density of the surrounding gas, and $\Gamma_{\rm jet}$ is the Lorentz factor of the jet. 

We discuss detectability of the emission from the jets for three models: model A assumes massive stellar mass BHs that corresponds to the system of GW150914 \citep{LIGO16a}, model B assumes intermediate mass BHs (IMBH) that are expected to exist in the center of star cluster \citep[e.g.,][]{GPP02a}, and model C assumes massive BHs (MBH) that may be formed by collapse of supermassive stars \citep[e.g.,][]{ST83a}, for which $\mbh$, $\rini$, and resultant physical quantities are tabulated in Table \ref{tab:EM}. The other parameters are fixed as $\alpha=0.1$, $\Tdead=3000$ K, $\asep=0.3$, $\Gamma_{\rm jet}=10$, and $n_{\rm ext}=1$ cm$^{-3}$. The durations of the jet launch and the afterglow are longer for higher $\mbh$ and larger $\rini$, and $L_{\rm AG}$ is proportional to $\mbh$. 

Figure \ref{fig:detect} shows time evolution of the internal shock emission flux in a certain energy band, $F_{\rm band}=\eta_{\rm band}F_\gamma$, for models A, B, and C with the values of $d_{\rm L}$. We set $\eta_{\rm band}\eta_\gamma\sim0.1$ for simplicity. The 1-scan sensitivity of Gas Slit Camera (GSC) on Monitor of All-sky X-ray Image (MAXI) for an energy range 2 keV--20 keV is also plotted \citep{MAXI16a}. Since it takes 40 s--150 s for the 1-scan of MAXI, these jets are detectable if $F_{\rm band}$ at $t_{\rm mer}-t=$ 40 s is higher than the sensitivity. We can see that the emissions are marginally detectable for all the models. Since the sensitivity of {\it Swift} Burst Alert Telescope (BAT) for 15 keV--150 keV with exposure time of 40 s is comparable to that of MAXI \citep{Swift05a}, these jets are detectable even if they mainly emit hard X-rays. This emission is a unique electromagnetic counterpart of GWs from merging BBHs in the sense that it can be detectable before the GW signal and that the luminosity increases with time. However, the luminosity and total energy are too low to explain GRBs or the GBM event \citep{FermiGBM16a}.

The optical follow-up surveys for GW counterparts, such as the Pan-STARRS1 and the Japanese collaboration for Gravitational wave ElectroMagnetic follow-up (J-GEM), have a sensitivity of 19--21 magnitude \citep{Pan-STAR16a,J-GEM16a}. Assuming 10 \% of $L_{\rm AG}$ is in the optical range, we calculate the distance of detection limit for the afterglow, $d_{\rm L,lim}$, whose values are tabulated in Table \ref{tab:EM}. While the afterglow is fainter than the internal shock emission, the optical follow-up surveys of afterglows can detect comparable or more distant events than the X-ray monitoring systems owing to the good sensitivity. However, $d_{\rm L,lim}$ for model A is shorter than the distance of the observed GW events \citep{LIGO16a,LIGO16d}. Although $d_{\rm L,lim}$ is larger for IMBHs and MBHs, we do not discuss the detection probability because the merger rates of IMBH binaries and MBH binaries are very uncertain.

\section{Summary \& Discussion}\label{sec:summary}

We study evolution of an accretion disc in BBH systems and propose an evolutionary track of the disc, which leads to different conclusion from the previous work \citep{PLG16a}. At first, the disc viscously expands outward but the companion prevents the disc from expandig beyond $r_{\rm out}$ due to the tidal torque. The evolution of viscous disc results in the decrease of the disc mass and the temperature. When the disc sufficiently cools down (typically less than 3000 K), the dead disc forms because MRI becomes inactive. Since the thermal instability causes the rapid drop of the disc temperature, the disk becomes dead when the temperature becomes less than a few tens of thousands K. This dead disc remains until the binary separation sufficiently decreases. As the binary separation decreases, the position at which the tidal torque is effective moves inward, and the mass of the outer rim increases. Then, the angular momentum is transported by the tidal torque, which induces the mass inflow from the outer rim to the dead disc. When the mass inflow by the tidal torque becomes higher than $\dot M_{\rm dead}$, the accretion heating activates MRI, restarting the mass accretion from the disc to the central black hole (the disc revival). This disc revival typically happens tens of thousands years before the merger event. The evolution of the revival disc is determined by the tidal torque, keeping $t_{\rm vis}\sim t_{\rm GW}$. The mass accretion rate of the revival disc increases with time.

In the late phase of the revival disc evolution, the mass accretion rate can exceed Eddington rate, and a relativistic jet is expected to be launched. We estimate the electromagnetic flux from the jet and discuss its detectability. Since the jet luminosity is increasing with time, the X-ray flux from the internal shock increases with time. This flux can be detectable before the merger event. The afterglow can typically be luminous a few thousands seconds after the merger. The estimated flux from the jet is too low to explain the GBM event, but detectable by the optical transient surveys or X-ray monitoring systems if the merger events happen in the local universe ($\lesssim10\rm~Mpc$) or if BHs are very massive ($\sim10^5~M_{\odot}$). 

In Section \ref{sec:initial}, the disc physical quantities is mildly inconsistent with the thin-disc approximation in the early phase. When $H/r>1$ and $p_{\rm rad}>p_{\rm gas}$, we should use the slim disc solution that has different features from the standard thin disc \citep{abr+88,CG09a}. In this regime, the disc mass decreases more rapidly than the standard thin disc, which shorten $t_{\rm dead}$. When the mass accretion rate becomes lower than the Eddington rate, the disc state changes from the slim disc to the thin and radiation-pressure dominant disc \citep{abr+88,KFM08a}. This regime is thermally unstable \citep{SS76a}. Some models with a different expression of the stress can avoid this instability \citep{SC81a,HBK09a}. However, the most recent simulation with a wide calculation range and a better radiative transfer scheme shows that the solution is thermally unstable \citep{jia+13}, and it is unlikely to be realized. Thus, the disc state is expected to change to the thin and gas-pressure dominant disc soon after the slim disc regime ends. Since $\dot M_{\rm dead}$ is much less than the Eddington rate, the thin and gas-pressure dominant disc takes place whenever the disc becomes dead. Therefore, even if we address the disc evolution discussed above, our estimate in Section \ref{sec:evolv} would not change except that $t_{\rm dead}$ would be shortened. The shortened $t_{\rm dead}$ could not affect our statement that the disc evolution time is much shorter than the decreasing time of binary separation.

In Section \ref{sec:dead}, we ignore ionization by cosmic rays (CRs), although its effect for accretion process is still under debate \citep[e.g.,][]{BS13a}. The CRs ionize the disc surface layer of $\Sigma_z=\int_z^\infty\rho(z)dz\lesssim100~\rm~g~cm^{-2}$, where $\rho(z)$ is the density \citep{UN81a}. Assuming the density of CRs is the same as that in the interstellar medium of the Galaxy, we write the ionization rate as $\zeta_{\rm cr}\sim10^{-17}\rm~cm^3~s^{-1}$ \citep{UN81a}. The equilibrium condition between the ionization by CRs and recombination is $\zeta_{\rm cr}n_H=\beta_{\rm rec}n_en_p$, where $\beta_{\rm rec}=6.22\times10^{-13}T_{3.5}^{-3/4}$ is the radiative recombination rate \citep[the UMIST database,][]{UMIST12}. Assuming $n_H=\Sigma/(2m_p H)$, $n_e=n_p$, and $n_e=\chi_e n_H$, we obtain the equilibrium ionization degree $\chi_{\rm cr}$. The instability condition for MRI is $\chi_{\rm dead}\le\chi_{\rm cr}$. We calculate the critical $\beta_{\rm pl}$ below which the MRI is active, whose values are 26, 2.2$\times10^2$, and $8.9\times10^3$ for models A, B, and C in Section \ref{sec:detect}, respectively. Since the expected value of $\beta_{\rm pl}$ by the MRI turbulence ranges 10--100, the layered accretion is likely for models B and C. In this case, the surface layer of $\Sigma_{\rm active}\sim 100\rm~g~cm^{-2}$ accretes onto BHs \citep{Gam96a}. The mass loss by the layered accretion in $t_{\rm mer}$ is estimated to be $M_{\rm lay}\sim3\pi\nu\Sigma_{\rm active}t_{\rm mer}$, the values of which are 46 $M_{\odot}$, 68 $M_{\odot}$, and 1.6$\times10^3M_{\odot}$ for models A, B, and C, respectively. These values are much higher than the mass of the dead disc. Therefore, some mechanism to reduce CR density is necessary to leave the dead disc until the disc revives for models B and C. 

In Section \ref{sec:revival}, we use some assumptions such as constant separation parameter ($\asep=0.3$) and $\widetilde{\Sigma}\sim m_{\rm dead}/(\pi r_{\rm out}^2) $. In order to verify these assumptions, we should perform long-term non-ideal magneto-hydrodynamical simulations with cold fluid and non-axisymmetric gravity.  This is because (a) the viscous time is much longer than the dynamical time, (b) the resistivity is essential for the death and revival of the disc, (c) the sound speed in the disc is much slower than the Keplerian velocity, and (d) tidal torque is non-axisymmetric effect. Such simulations remain as a future work.

\section*{Acknowledgements}
We thank Takanori Sakamoto, Kohta Murase, and Yuri Fujii for useful comments. This work is partly supported by JST grant ``Building of Consortia for the Development of Human Resources in Science and Technology'' (S.S.K. and K.T.), by JSPS Grants-in-Aid for Scientific Research 15H05437 (K.T.), by NASA NNX13AH50G (S.S.K), and by IGC post-doctoral fellowship program (S.S.K).

\bibliographystyle{mnras}
\bibliography{astro}

\bsp	
\label{lastpage}
\end{document}